\documentclass[amsmath,amssymb,aps,pra,twocolumn, superscriptaddress,longbibliography]{revtex4-2}

\usepackage{graphicx}
\usepackage{dcolumn}
\usepackage{xcolor}

\usepackage{helvet}
\usepackage{mathrsfs}
\usepackage[colorlinks=true,bookmarks=false,citecolor=blue,urlcolor=blue]{hyperref}
\usepackage{physics}
\usepackage{dsfont}

\renewcommand{\cite}[1]{[\onlinecite{#1}]}

\DeclareMathOperator{\cn}{cn}
\DeclareMathOperator{\sn}{sn}
\DeclareMathOperator{\dn}{dn}

\begin{document}

\title{Quantum advantage in transfer of  quantum states
}

\author{Andrei~A.~Stepanenko}
\email{as@lims.ac.uk}
\affiliation{London Institute for Mathematical Sciences, Royal Institution, London, UK}
\affiliation{School of Physics and Engineering, ITMO University, Saint Petersburg 197101, Russia}

\author{Kseniia S. Chernova}
\affiliation{School of Physics and Engineering, ITMO University, Saint Petersburg 197101, Russia}

\author{Maxim~A.~Gorlach}
\email{m.gorlach@metalab.ifmo.ru}
\affiliation{School of Physics and Engineering, ITMO University, Saint Petersburg 197101, Russia}

\begin{abstract}
Quantum advantage broadly understood as the ability of quantum systems to significantly outperform their classical counterparts underpins current interest to quantum technologies and is a topic of active investigation. In many situations, its existence is subject to debate and the areas of supremacy of large-scale quantum systems are not well defined. Here, we uncover a novel niche where quantum advantage can be clearly defined and proven. We study a time-optimal transfer of excitations in the lattice involving both nearest-neighbor and longer-range couplings. We prove that the quantum-mechanical property of a particle to propagate along several trajectories simultaneously speeds up the transfer process which takes shorter time compared to any particular trajectory and thus provides a clear example of quantum advantage.


\end{abstract}

\maketitle

\begin{figure}
    \centering
    \includegraphics[width=1\linewidth]{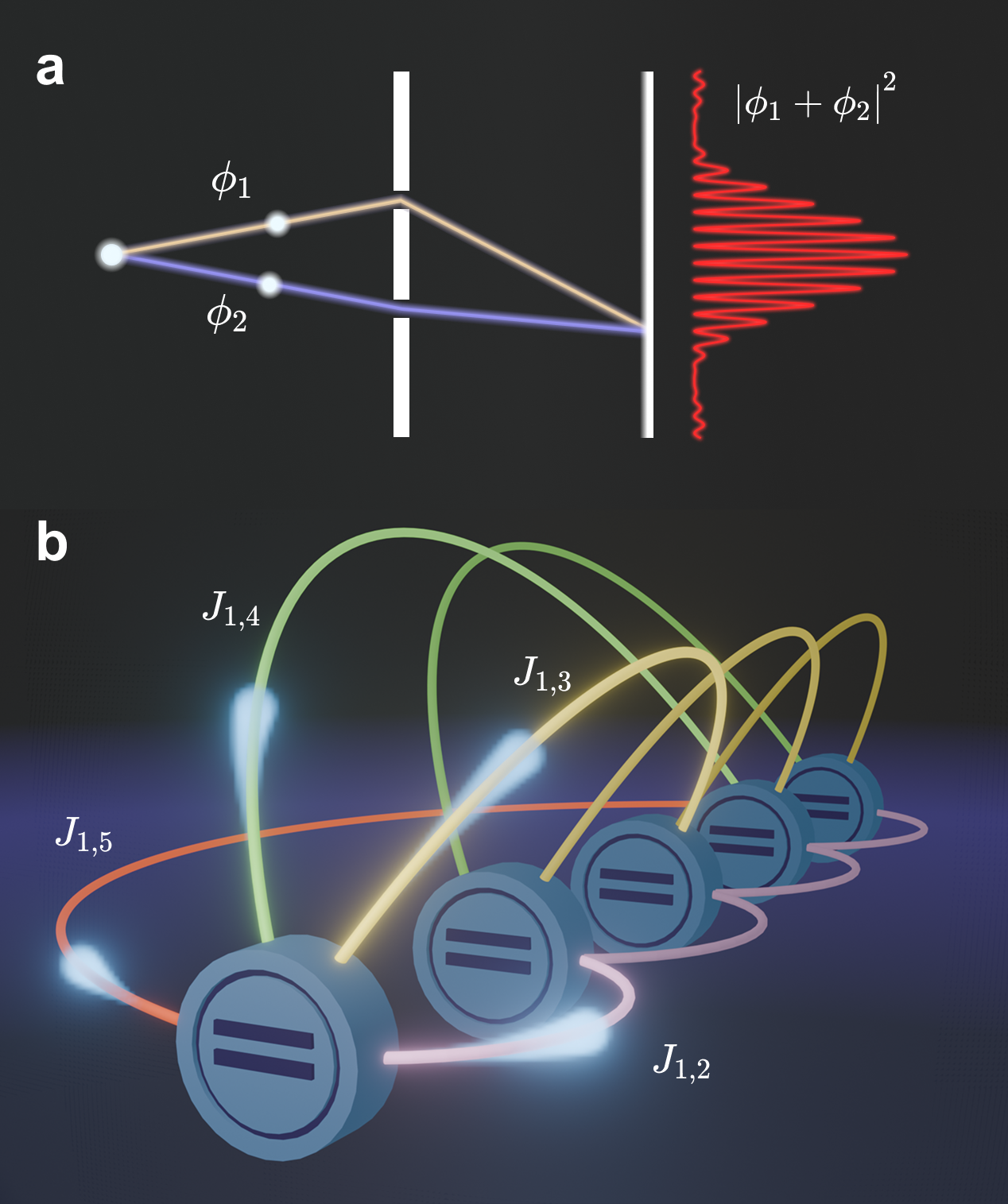}
    \caption{{\bf Illustration of quantum interference and its applications in quantum state transfer.} (a) Double slit experiment: an electron takes two paths simultaneously producing an interference pattern at the screen. 
    (b) A lattice of $N$ qubits with the nearest-neighbor and longer-range couplings. All  couplings $J_{m,n}$ between the qubits $m$ and $n$ are controlled in real time.
    }
    \label{fig:1}
\end{figure}

Recent progress in quantum technologies has been largely driven by the pursuit of {\it quantum advantage}~--~the ability of quantum devices to solve certain tasks more efficiently than their classical counterparts~\cite{Har17,Rnn14,Boi16,Aar10,Han22}. A major direction is the computational quantum advantage~\cite{Har17,Aar10,Han22} demonstrated for quantum computing tasks in superconducting~\cite{Arute2019,JianWeiPan2021,Gao2025} and photonic~\cite{ZhongPan2020,Madsen2022,Oh2024} architectures. A distinct area is quantum illumination, which promises better target detection in noisy and lossy environments by utilizing quantum sources of light~\cite{Tan2008,Lloyd2008,Lopaeva2013}. A recent concept is quantum batteries where quantum entanglement accelerates the charging process~\cite{Binder2015,Campaioli2017,Hu2026}. These and other applications including quantum annealing~\cite{Bauza2025} and analog quantum simulation~\cite{King2025} demonstrate the advantages of quantum devices for solving specific tasks. However, a  broader picture is currently lacking.


At the same time, it is understood that existing demonstrations of quantum advantage largely build on such foundational concepts of quantum physics as superposition, interference~\cite{Gro97,Stahlke2014}, and entanglement~\cite{Jozsa2003,Vidal2003}. A proper combination of these ingredients allows one to coherently interfere the probability amplitudes, amplifying the desired outcomes while suppressing others and achieving massive parallelism~\cite{Gro97,Stahlke2014}.



While the interference phenomenon is ubiquitous for classical waves, quantum interference is more subtle and occurs even for a single particle. This was articulated in the famous double-slit thought experiment by Feynman~\cite{Feynman2010}. A single electron incident at the screen with the two slits can take any of the two paths [Fig.~\ref{fig:1}{\bf a}]. From the classical perspective, one expects the electron to pass either through the first or through the second slit, producing two Gaussian-shaped probability distributions at the output screen. However, in the quantum case, the electron takes both paths {\it simultaneously} which results in the distinctive interference pattern at the output. This striking feature of quantum physics has been tested experimentally~\cite{Juffmann2012, Bach2013}
and nowadays provides a compelling demonstration of the laws of quantum physics. 

Here, we demonstrate that this distinctive feature of quantum mechanics enables useful functionality in the celebrated problem of quantum state transfer~\cite{Christandl2004,Christandl2026}, allowing to speed up the transport of excitations in qubit lattices and revealing yet another facet of quantum advantage.

We consider a one-dimensional (1D) array of identical qubits involving nearest-neighbor and longer-range couplings which can be reconfigured in real time [Fig.~\ref{fig:1}{\bf b}], which is technically feasible for existing qubit lattices~\cite{Yan2018}. We assume that the single-particle excitation is launched in the first qubit of the array and seek the fastest way to transfer it to the rightmost $N^{\text{th}}$ qubit. Due to the physical constraints, the maximal value of the long-range couplings achievable in such a setup is suppressed compared to the shorter-range interaction. This physical limitation is reflected in the constraint on the system Hamiltonian:
\begin{equation}\label{eq:constraint}
\sum\limits_{p=1}^{N-1}\,g_p\,\sum\limits_{m=1}^{N-p}\,J_{m,m+p}^2(t)=J_0^2\:.
\end{equation}
Here $J_{m,m+p}$ is a real-valued time-dependent coupling between the sites $m$ and $m+p$, $J_0$ is the overall coupling resource assumed to be constant, while $g_p$ are weights increasing with $p$ and describing the suppression of the long-range interactions with the distance $p$ between the sites.

Given physically motivated constraint Eq.~\eqref{eq:constraint}, we seek the optimal way to vary the couplings $J_{m,m+p}$ in time to achieve the fastest possible transfer with the fidelity equal to 1. Naively, one may imagine a strategy when the particle starts in the first qubit, hops from site to site  and reaches eventually $N^{\text{th}}$ qubit. In analogy to the Feynman's double-slit experiment, we call this scenario a {\it classical trajectory}. If the particle does not bounce back and forth in the lattice (which is clearly far from the optimum), the number of classical trajectories is $2^{N-2}$, where $N$ is the number of qubits.

However, the excitation does not have to follow any particular classical trajectory: quantum mechanics allows it to take several or many paths simultaneously. Below, we show that if the long-range interactions are sufficiently constrained, this quantum-mechanical feature leads to a sizable and observable speedup in the transfer process providing a clear instance of quantum advantage. The Hamiltonian of the qubit lattice under study reads:
\begin{align}  &\hat{H}=\sum_{m=1}^{N-1}\sum_{p=1}^{N-m}J_{m,m+p}\left(i^{p-1}\ket{m}\bra{m+p}\right.\notag\\
    &+\left.i^{1-p}\ket{m+p}\bra{m}\right)+\omega_0\,\sum\limits_{m=1}^{N}\,\ket{m}\bra{m}\:,
\end{align}
where $\hbar=1$, $J_{m,n}$ is the real coupling amplitude between the sites $m$ and $n$ and $\omega_0$ is the eigenfrequency of qubits which defines the reference energy and is omitted further for brevity. The effects of dissipation are neglected which is a valid assumption for state-of-the-art qubit lattices~\cite{Somoroff2023}. The vectors $\ket{m}$ form the basis in the $N$-dimensional single-particle sector of the Hilbert space. Importantly, the hoppings $J_{m,m+p}$ contain the phase factors  $i^{p-1}$ and $i^{1-p}$ which are crucial to ensure the constructive interference of the different propagation paths.

To demonstrate quantum advantage in transfer of the quantum states, we employ quantum brachistochrone technique~-- a variational method aiming to minimize time needed for transition from the given initial to the prescribed final state for the specified constraints [Eq.~\eqref{eq:constraint}] on the system Hamiltonian~\cite{Carlini2006,Carlini2007,Wang2015,Koike2022}. This machinery can be viewed as a particular case of Pontryagin maximum principle~\cite{Boscain2021,Koike2022} and was previously applied to optimize small-scale systems~\cite{Carlini2011,Carlini2012,ZhenWang2025,Chernova2025,Meinersen2024,WangeSong2025,WangeSongPRL2025},
but recently has been significantly upgraded to describe arrays with hundreds of qubits~\cite{Stepanenko2025} and, potentially, even infinitely large lattices~\cite{stepanenko2025arxiv} providing an efficient alternative to numerical optimization schemes~\cite{Lukin2025}.

The dynamics of a quantum state $\ket{\psi(t)}$ is encoded in the unitary evolution operator defined via $\ket{\psi(t)} = \hat{U}(t)\ket{\psi(0)}$ and evolving according to the Schr{\"o}dinger equation $i\,\partial_t\hat{U} = \hat{H}\hat{U}$. To find the optimal transfer protocol, we minimize the cost functional
\begin{align}
    \label{eq:action}
    S[\hat{U},\lambda_s,\hat{R}] =& \frac{1}{2}\,\int_0^\tau\,\Tr(\hat{G}(\hat{H})\hat{H})\,dt+\int_0^\tau\,\Tr(\hat{D}\hat{H})\,dt\notag\\
    + &\int_0^\tau\,\Tr(\hat{R}\,(\partial_t\hat{\rho}+i[\hat{H},\hat{\rho}])))dt
\end{align}
assuming $\delta\hat{\rho}(0)=\delta\hat{\rho}(\tau)=0$ due to the specified initial and final quantum states. The operator $\hat{G}(\hat{H})$ incorporates additional penalties $g_p$ on the long-range couplings:
\begin{align}
& \hat{G}(\hat{H})=\sum_{m=1}^{N-1}\sum_{p=1}^{N-m}g_p\,J_{m,m+p}\left(i^{p-1}\ket{m}\bra{m+p}\right.\notag\\
    &+\left.i^{1-p}\ket{m+p}\bra{m}\right)\:.    
\end{align}
Thus, the first term in Eq.~\eqref{eq:action} takes into account constraint Eq.~\eqref{eq:constraint} and minimizes the resource $J_0$ needed to transfer the quantum state within fixed time $\tau$ which is equivalent to minimizing transfer time for the fixed resource $J_0$. 

The second term constrains the form of the Hamiltonian via the operator $\hat{D}=\sum_{s}\lambda_{s}\hat{D}_s$ with time-dependent Lagrange multipliers $\lambda_s$. Variation with respect to $\lambda_s$ yields the set of conditions $\text{Tr}\,(\hat{H}\,\hat{D}_s)=0$ at all moments of time. For instance, this excludes time-varying qubit eigenfrequencies and ensures the phase of short- and long-range couplings necessary for the interference.



Finally, the third term in Eq.~\eqref{eq:action} guarantees that the density matrix $\hat{\rho} = \ket{\psi}\bra{\psi}$ evolves according to von Neumann equation obtained by varying the functional with respect to $\hat{R}$. As a consequence, the density matrix depends on time as $\hat{\rho}(t)= \hat{U}(t)\,\hat{\rho}_0\,\hat{U}^\dagger(t)$ and $\hat{R}(t)$ is an unknown Hermitian time-dependent Lagrange multiplier matrix.

Varying the cost functional with respect to $\hat{U}$ and $\hat{\rho}$, we obtain quantum brachistochrone equation (QBE)~\cite{Carlini2006, Carlini2007,Wang2015,Koike2022} with the boundary conditions~\cite{Stepanenko2025}
%
\begin{eqnarray}
    \label{eq:QBE}
    \partial_t\hat{L} &=& -i[\hat{H},\hat{L}]\,,\\
    \label{eq:QBE_b1}
    \hat{L}(0) &=& i[\hat{R}_0,\hat{\rho}_0] \,,\\
    \label{eq:QBE_b2}
    \hat{L}(\tau)&=& i[\hat{R}_\tau,\hat{\rho}_\tau]\;,
\end{eqnarray}
which govern the evolution of the Hamiltonian (see Supplementary Section 1). Here $\hat{L}= \hat{G}(\hat{H})+\hat{D}$ and $\hat{\rho}=\ket{\psi}\bra{\psi}$. These equations should be solved jointly with the Schr{\"o}dinger equation defining the evolution of the wave function
\begin{equation}\label{eq:Schr}
i\,\partial_t \ket{\psi}=\hat{H} \ket{\psi}\:.
\end{equation}
Equations~\eqref{eq:QBE_b1}-\eqref{eq:QBE_b2} hold not only at $t=0$ and $t=\tau$, but also at the arbitrary moment of time, suggesting a synchronized change of the Hamiltonian and the wave function.

The analysis of the problem is simplified by observing that the $(N^2-1)$-dimensional space of traceless Hermitian $N\times N$ matrices splits into two non-overlapping parts: subspace $\mathcal{A}$ whose elements $\hat{A}$ anticommute with the chiral symmetry operator $\hat{\Sigma}$:  $\hat{A}\,\hat{\Sigma}+\hat{\Sigma}\hat{A}=0$ and the remaining subspace $\mathcal{D}$ whose elements commute with $\hat{\Sigma}$. Here, chiral symmetry operator is defined via $\hat{\Sigma}=\hat{Q}^2\,\hat{K}$, where $\hat{K}$ is complex conjugation and $\hat{Q} = \sum_m i^{m-1}\ket{m}\bra{m}$. As a result, the dynamics of the two parts of $\hat{L}$ operator~-- $\hat{L}_A\in\mathcal{A}$ and $\hat{L}_D\in\mathcal{D}$~-- gets decoupled, so that $\hat{L}_A=\hat{G}(\hat{H})$ satisfies the same Eqs.~\eqref{eq:QBE}-\eqref{eq:QBE_b2} (Supplementary Section 1). 

Even stronger simplification comes from the algebraic structure of Eqs.~\eqref{eq:QBE}-\eqref{eq:Schr} indicating that $\hat{L}_A$ and $-i\hat{H}$ form a Lax pair~\cite{stepanenko2025arxiv} revealing an integrable nature of quantum brachistochrone problem~\cite{Malikis2024}. As a result, the eigenvalues of the Lax operator $\hat{L}_A$ do not depend on time. In turn, the boundary conditions Eqs.~\eqref{eq:QBE_b1}-\eqref{eq:QBE_b2} ensure that only two of those eigenvalues are nonzero, having the same magnitude $L_0$ and the opposite signs. Taken together, this yields
\begin{equation}\label{eq:LaxExpr}
    \hat{G}(\hat{H})\equiv\hat{L}_A = L_0(\ket{a}\bra{a}-\ket{b}\bra{b})\;,
\end{equation}
where $\pm L_0$ are two unknown non-zero eigenvalues, and the vectors $\ket{a}$ and $\ket{b}$ related via $\ket{b}=\hat{\Sigma}\,\ket{a}$ evolve according to Eq.~\eqref{eq:Schr}. At the same time, the wave function is expressed through the same vectors as
\begin{equation}\label{eq:WExpr}
\ket{\psi}=\frac{1}{\sqrt{2}}\,\left(\ket{a}+\ket{b}\right)\:.
\end{equation}
As a consequence, the complexity of the optimal control problem is greatly reduced, as both the Hamiltonian and the wave function with $\sim N^2$ and $N$ entries, respectively, are expressed via single auxiliary $N$-component vector.

To simplify the notations, we perform a basis transformation $\ket{\alpha}=\hat{Q}\ket{a}$, $\ket{\beta}=\hat{Q}\ket{b}$ and $\hat{H}^Q=\hat{Q}\hat{H}\hat{Q}^{-1}$. In such a basis, chiral symmetry reduces to the complex conjugation,
\begin{align}
& \ket{\psi^Q}=\left(\ket{\alpha}+\ket{\alpha^*}\right)/\sqrt{2}\:,\label{eq:psiviaa}\\
& J_{m,n} = iL_0\dfrac{\alpha_m \alpha_{n}^* - \alpha_m^* \alpha_{n}}{g_{|m-n|}}\:,\label{eq:Jviaa}
\end{align}
and the governing equation for $\alpha_m$ amplitudes takes the form
\begin{eqnarray}
    \label{eq:sys_a}
    i\partial_t \alpha_m &=& L_0\sum_{n=1}^{N}\dfrac{\alpha_m \alpha_{n}^* - \alpha_m^* \alpha_{n}}{g_{|m-n|}} \alpha_n\;.
\end{eqnarray}
The boundary conditions read 
%
\begin{eqnarray}
    \label{eq:ba1}
    \Re(\alpha_m(0)) = \delta_{m1}/\sqrt{2}\;,\\
    \label{eq:ba2}
    \Re(\alpha_m(\tau)) = \delta_{mN}/\sqrt{2}\;.
\end{eqnarray}
The problem Eqs.~\eqref{eq:sys_a}-\eqref{eq:ba2} is solved by the shooting method which requiers a good initial approximation for $\Im(\alpha_m(0))$ to converge. Starting from small systems, where the correct initial guess can be found directly, we extrapolate the initial vector $\ket{\alpha(0)}$ based on the data for the smaller system and assuming only a minor modification of its structure with the increase of $N$ (Supplementary Section 2).

\begin{figure}[b]
    \centering
    \includegraphics[width=1\linewidth]{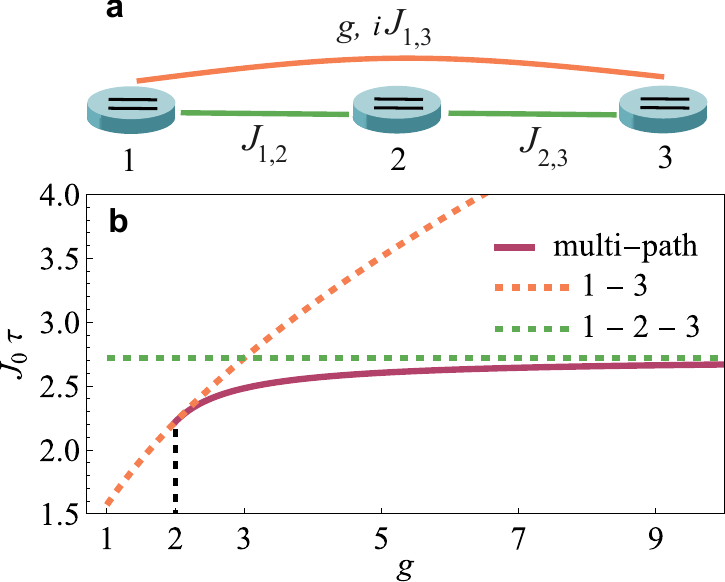}
    \caption{{\bf Speedup of state transfer due to quantum interference.} (a) Scheme of the three-qubit array. Complex long-range coupling $iJ_{1,3}$ is suppressed by the weight $g$ compared to the short-range couplings $J_{1,2}$ and $J_{2,3}$. (b) Time of transfer from the first to the third qubit as a function of weight $g$. The optimal scenario (magenta solid line) is faster than consecutive 1-2-3 hopping (green dashed line) or direct 1-3 transfer (orange dashed line) and involves superposition of both paths. This solution exists provided $g\geq2$.}
    \label{fig:2}
\end{figure}


After establishing a general methodology, we illustrate the concept of quantum advantage on a simple but instructive example. We consider a lattice of 3 qubits where an imaginary reciprocal coupling $iJ_{1,3}$ between spatially distant the first and the third qubits is suppressed by the weight $g$ [Fig.~\ref{fig:2}{\bf a}], i.e. the constraint on the couplings has the following form:
\begin{equation}\label{eq:3qconstraint}
J_{1,2}^2+J_{2,3}^2+g\,J_{1,3}^2=J_0^2\:.
\end{equation}
Under this constraint, we seek the fastest protocol transferring the excitation from the first to the third qubit. There are two routes for the excitation: it can either jump from the first to the last qubit directly, or hop through the qubits consecutively.

If the excitation takes the direct $1-3$ route, $J_{1,3}$ coupling should be maximal and equal to $J_0/\sqrt{g}$. The time of the transfer is readily evaluated as
\begin{equation}
\tau_d=\frac{\pi\,\sqrt{g}}{2\,J_0}\:.
\end{equation}
This scenario is depicted in Fig.~\ref{fig:2}{\bf b} by the orange dashed line.

On the other hand, the excitation may take an alternative $1-2-3$ path hopping consecutively  between the qubits. In such case with $J_{1,3}=0$, the fastest option is to vary $J_{1,2}$ and $J_{2,3}$ couplings as cosine and sine functions of time, which results in the time of the transfer~\cite{Chernova2025}
\begin{equation}
\tau_{n} = \frac{\sqrt{3}\pi}{2J_0}\:,
\end{equation}
which does not depend on the weight $g$. This scenario is illustrated in Fig.~\ref{fig:2}{\bf b} by the green dashed line.

An interesting question is which path wins the competition and whether the protocol combining both trajectories simultaneously could provide an advantage. To address that, we apply quantum brachistochrone equation~\eqref{eq:QBE} and the boundary conditions Eqs.~\eqref{eq:QBE_b1}, \eqref{eq:QBE_b2} which read
%
\begin{align}
    &\partial_t J_{1,2} = -(g-1)J_{2,3}J_{1,3}\:,\\
    &\partial_tJ_{2,3} = (g-1)J_{1,2}J_{1,3}\:,\\
    &\partial_tJ_{1,3} = 0\:,\\
    &J_{2,3}(0)=0\:,\mspace{6mu} J_{1,2}(\tau)=0\:.
\end{align}
Solving above equations jointly with the Schr{\"o}dinger equation Eq.~\eqref{eq:Schr} and the conditions for the wave function $\ket{\psi(0)} = (1, 0, 0)^T$ and $\ket{\psi(\tau)} = (0, 0, -1)^T$ (see Supplementary Section 4), we recover:
\begin{eqnarray}
    J_{1,2}(t) &=& J_0\,\sqrt{\dfrac{(g-2)(3g-2)}{(g-1)(3g-4)}}\cos(\Omega\, t)\;,\label{eq:3qJ1}\\
    J_{2,3}(t) &=& J_0\,\sqrt{\dfrac{(g-2)(3g-2)}{(g-1)(3g-4)}}\sin(\Omega\, t)\;,\label{eq:3qJ2}\\
    J_{1,3}(t) &=& J_0/\sqrt{(g-1)(3g-4)}\;,\label{eq:3qJ3}\\
    \Omega &=& J_0\,\sqrt{\dfrac{g-1}{3g-4}}\:,\label{3q:omega}
\end{eqnarray}
where $g\geq2$. The time of the transfer thus reads:
\begin{equation}\label{eq:Tau3q}
\tau=\dfrac{\pi}{2J_0}\sqrt{\dfrac{3g-4}{g-1}}\:,
\end{equation}
and its dependence on the weight $g$ is shown in Fig.~\ref{fig:2}{\bf b} by the solid magenta line. Equations~\eqref{eq:3qJ1}-\eqref{eq:3qJ3} suggest that the optimal protocol utilizes both propagation paths simultaneously, while Fig.~\ref{fig:2}{\bf b} indicates that the protocol is faster than {\it any} of the two classical trajectories. For instance, for $g=3$ any of the two classical routes takes the same time $\tau_{n}=\pi\sqrt{3}/(2J_0)$. However, harnessing them simultaneously reduces the time of the transfer down to $\tau\approx 0.91\,\tau_{n}$.

A caveat to this solution is that the longer-range coupling $J_3$ must be sufficiently constrained, and the respective weight $g$ should be larger than 2. Otherwise, the fastest route is simply the direct hopping from qubit 1 to qubit 3 (Supplementary Section 3).

On the other hand, large weights $g$ restrict the coupling $J_{1,3}$ too strongly, and then the optimal protocol approaches asymptotically the scenario with the consecutive hopping between the qubits, so that quantum interference effects are not pronounced.

While the gains from the quantum interference look modest in the three-qubit case, the situation changes dramatically if the array size $N$ is increased. The number of classical trajectories connecting the first and the last qubits grows exponentially as $2^{N-2}$ (Supplementary Section~5), and their interference promises a significant speedup.

To test that, we examine a lattice of qubits with all-to-all connectivity and the weights $g_p=p^2$, so that the maximal coupling between $m$ and $m+p$ sites is inversely proportional to the distance between them, which is the case for some trapped ion chains~\cite{Britton2012,Nevado2017}. Under this assumption, we calculate the minimal transfer time for all classical trajectories in the lattices with 3,4,5 and 6 qubits. The results are shown by the yellow dots in the inset of Fig.~\ref{fig:3} (see data in Supplementary Table 2).

As the length $N$ of the lattice increases further, the analysis of all $2^{N-2}$ classical trajectories becomes prohibitively hard, while the respective sets of transfer times form a continuum shown in Fig.~\ref{fig:3} by light blue color.

Despite the complexity of the problem, the smallest transfer time for the entire set of classical trajectories at given $N$ can be reliably estimated. As it is known~\cite{stepanenko2025arxiv}, in the absence of the boundary effects the optimal transfer of excitation via consecutive hopping between the nearest-neighbor coupled qubits takes time
\begin{equation}\label{eq:classicaltr}
    J_0\,\tau_{cl}(N)=1.13031\,(N-1)\:.    
\end{equation}
In our lattice with all-to-all connectivity, the excitation can hop by longer distances, e.g., between the sites $m$ and $m+p$. However, the gain in the distance is balanced by the decrease in the speed of the transfer, which is proportional to $J_{m,m+p}\propto1/\sqrt{g_p}=1/p$. Therefore, up to the boundary effects, the trajectory with the longer-range couplings takes the same time, Eq.~\eqref{eq:classicaltr}. Reshaping of the wave packet close to the boundaries or during the propagation in the lattice takes extra time, and hence Eq.~\eqref{eq:classicaltr} provides a lower bound for the entire family of classical trajectories as discussed further in the Supplementary Section 5. Comparison of this estimate to the calculated transfer times for short arrays ($3\leq N\leq6)$ indicates that Eq.~\eqref{eq:classicaltr} is overly optimistic, and the actual time of transfer for all classical trajectories is larger. 

\begin{figure}[t]
    \centering
    \includegraphics[width=1\linewidth]{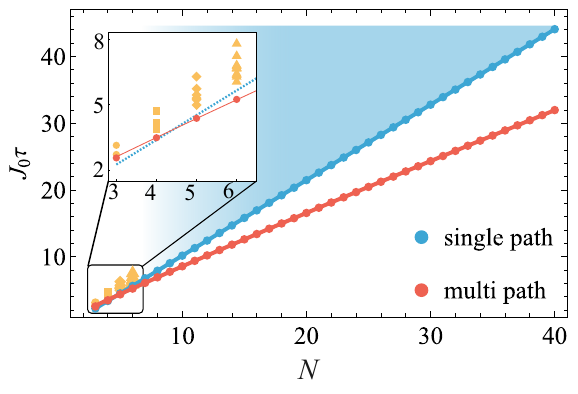}
    \caption{{\bf Demonstration of quantum advantage in transfer of quantum states.} 
    The transfer times for the optimal protocol shown by the red dots are well below the estimate Eq.~\eqref{eq:classicaltr} for any classical trajectory shown by blue. The latter, in turn, is below the actual transfer times for classical trajectories shown by the yellow markers and computed for the relatively small arrays with 3,4,5 and 6 qubits.}
    \label{fig:3}
\end{figure}

Essentially, Eq.~\eqref{eq:classicaltr} provides an analog of Bell's inequality for transfer problems: if the transfer protocol features the transfer time $\tau<\tau_{cl}$, the transport is non-classical in the sense that it cannot be viewed as a sequence of hoppings of a localized excitation.

Next, we compute a time-optimal solution for the transfer in our lattice. The calculations suggest that the optimal protocol activates most of the couplings in the lattice, so that the probability currents flow via many paths simultaneously (Supplementary Figure S1). Thus, the interference of the different classical trajectories takes place. The calculated results for the transfer time are shown by the red dots in Fig.~\ref{fig:3}. Remarkably, the transfer time is {\it lower} than the estimate Eq.~\eqref{eq:classicaltr}, which means that the transport happens faster than {\it any} classical trajectory takes. This confirms the conclusion that the transport is mediated by quantum interference and is essentially non-classical. This demonstrates the concept of quantum advantage in transfer of quantum states.

For the studied lattices with $N\leq 40$, the transfer time scales sub-linearly $J_0\tau \sim N^z$ ($0<z\leq 1$) with the length of the array. 
However, with the increase of the array size $N$, the dominant part of the excitation concentrates in a finite set of sites with the negligible population in the rest of the array. Therefore, in the limit of large $N$ we anticipate a linear dependence $\tau(N)$.
We fit this asymptotic dependence by the formula
\begin{equation}
    J_0\tau = 0.757 \,N + 2.018\:,
\end{equation}
which provides approximately 33\% speedup compared to consecutive hopping (see Supplementary Section~6).

In summary, we have proved that the effect of quantum interference allows one to speedup the transfer of quantum states in the lattices of qubits with all-to-all connectivity. This uncovers a novel facet in the concept of quantum advantage, revealing a fundamental connection to the celebrated double-slit interference experiment.

We anticipate that more substantial advantage can be attained, when the initial or target quantum states are spatially extended which could enable even more efficient interference of all available classical trajectories.

These results also shine new light on the concept of computational quantum advantage, since the process of computation can be viewed as a sequence of quantum state preparation and measurement.




\bibliography{ref}

\vspace{30pt}

{\bf Methods}

{\bf Numerical solution of the QBE}

Equation~\eqref{eq:sys_a} for the dynamics of the Lax eigenvector is solved numerically. For efficient numerical computation, we separate real and imaginary parts $\alpha_m = x_m + i y_m$ and, accordingly:
\begin{eqnarray}
    \partial_t x_m &=& \sum_n \dfrac{2L_0}{g_{m,n}}(y_{m}x_n - x_{m}y_n) x_n\;,\label{eq:dynamics1}\\
    \partial_t y_m &=& \sum_n \dfrac{2L_0}{g_{m,n}}(y_{m}x_n - x_{m}y_n) y_n\label{eq:dynamics2}
\end{eqnarray} 
with the boundary conditions
\begin{align}
& x_m(0)=\delta_{m1}/\sqrt{2}\mspace{8mu} \text{for}\: 1\leq m\leq N,\\
& y_1(0)=0\:,\\
& x_m(\tau)=0 \mspace{8mu}\text{for}\: 1\leq m \leq N-1.
\end{align}
%
The system of $2N$ nonlinear  differential equations Eqs.~\eqref{eq:dynamics1}-\eqref{eq:dynamics2} with $2N$ boundary conditions is supplemented by the requirement 
\begin{equation}
\partial_t L_0 = 0\:.    
\end{equation}
Since the vector $\ket{a}$ has a constant norm $\left<a|a\right>=1$, a boundary condition  $\sum_m (x_m^2 + y_m^2) = 1$ is required.

This yields the system with a total of $2N+1$ variables, equations, and boundary conditions that can be solved by the shooting method or via gradient-based optimization. To converge, both of these methods require a sufficiently good initial guess for $\ket{y(0)}$ and $L_0$.  

For a small chain size $N<5$, the solution is found after a few attempts with random guesses. Based on the visual structure of the found solution, we approximate the initial guess by $y_m(0) = -0.05 + 1/\sqrt{2 (N-m+1)}$, for $2 \leq m\leq N$, which is not precise but simple enough and allows us to obtain results for lattice size $N<10$. In turn, the initial guess for $L_0$ is fitted by the $L_0 = 0.5\,N^{1.6} + 0.9 $ with the error below $0.5$ and can be extended for larger sizes. 

However, we were not able to find an analytical function for $y_m(0)$ providing a good initial approximation for larger arrays. To overcome this problem and calculate a transfer protocol in large chains, we use all already found $y_m^{(n)}(0)$, where the upper index corresponds to the size of the lattice $3\leq n\leq N_{max}$ and $N_{max}$ is the maximal size of a chain with a known solution. We now introduce the recurrent extrapolation procedure. 
\begin{enumerate}
    \item Interpolate the $y_m^{(n)}(0)$ by a smooth function $f^{(n)}(q)$,
    for all $3\leq n \leq N_{max}$ and $q\in[2, n]$.\\
    \item Discretize $f^{(n)}(q)$ to a common representation $f_z^{(n)} = f^{(n)}(z)$ consisting of $N_z = 25$ equally distributed points $1\leq z\leq N_z$.
    \item Interpolate $f_z^{(n)}$ by a smooth functions $F_z(n)$ of the size $n\in[3,N_{max}]$ for all $1\leq z\leq N_z$.
    \item Extrapolate $F_z(n)$ to $F_z(N_{max}+1)$ for $1\leq z\leq N_z$, assuming only a minor change in the initial guess and smoothness of $F_z(n)$.
    \item Define $f_z^{(N_{max}+1)} = F_z(N_{max}+1)$ and interpolate it by $f^{(N_{max}+1)}(z)$.
    \item Discretize $f^{(N_{max}+1)}(z)$ to $(N_{max}+1)$ equally distributed points $y_m^{(N_{max}+1)}(0) = f^{(N_{max}+1)}(m)$.
\end{enumerate}

The vector $y_m^{(N_{max}+1)}(0)$ serves as the initial guess for $y_m(0)$ in the shooting method. When a solution is found, we increase $N_{max}$ by one and repeat the procedure.


\vspace{10pt}

{\bf Data availability}

Data that support the findings of this study are available within the article or Supplementary Information. Source data are provided with this paper.

\vspace{10pt}

{\bf Code availability}

The code used for the data analysis in this study is available from the corresponding author upon reasonable request.

\vspace{10pt}

{\bf Acknowledgments}

We acknowledge Alexander Mikhalychev for valuable discussions. Theoretical models were supported by the Russian Science Foundation, grant No.~25-42-10012. Numerical simulations were supported by Priority 2030 Federal Academic Leadership Program.

\vspace{10pt}

{\bf Author contributions}

M.A.G. conceived the idea and supervised the project. A.A.S. and M.A.G. elaborated the theoretical methodology. K.S.C. and M.A.G. analyzed the three-qubit example. A.A.S. and K.S.C. performed numerical simulations and analyzed the data. All authors contributed to manuscript preparation.

\vspace{10pt}

{\bf Competing interests}
The authors declare no competing interests.


\clearpage 
\onecolumngrid
\begin{center}
    \bf{Supplementary Materials:\\ Quantum advantage in transfer of  quantum states}
\end{center}
\setcounter{section}{0} 
\setcounter{equation}{0} 
\setcounter{page}{1}
\renewcommand{\theequation}{S\arabic{equation}} 
\renewcommand{\thefigure}{S\arabic{figure}} 



\section{Optimal control}

In this section, we derive optimal control equations employing quantum brachistochrone technique~\cite{Carlini2006,Carlini2007,Wang2015,Koike2022} supplemented by the appropriate boundary conditions~\cite{Stepanenko2025} and tailored to efficiently simulate large-scale quantum systems~\cite{stepanenko2025arxiv}.

We consider a vector $\ket{\psi(t)} = \sum_m\psi_m(t)\ket{m}$ from the $N$-dimensional single-particle sector of the Hilbert space with the basis vectors $\ket{m}$ describing an excitation localized at site $m$. A unitary evolution from the initial state $\ket{\psi(0)}=\ket{\psi_0}$ to the target state $\ket{\psi(\tau)}=\hat{U}(\tau)\ket{\psi_0} = \ket{\psi_1}$ is governed by the Schr{\"o}dinger equation $i\partial_t\hat{U}=\hat{H}\hat{U}$, where Hamiltonian $\hat{H}(t)$ is $N\times N$ traceless Hermitian matrix. 

Assuming that the norm (that will be defined later) of the Hamiltonian 
is bounded $\|\hat{H}\| \leq J_0$, we rescale time $t\rightarrow \mu(t)$ by the arbitrary monotonically increasing function $\mu(t)$ with $|\mu'(t)|<1$, leading to the renormalization of the Hamiltonian $\hat{H}\rightarrow \hat{H}/\partial_t\mu(t)$ without the change of the evolution operator $\hat{U}(t)$. $\mu(t)$ function is chosen such that the renormalized Hamiltonian has a norm equal to $J_0$. Due to $\mu'(t)<1$, this rescaling leads to the speedup of the evolution. We therefore conclude that the minimal evolution time is attained for the Hamiltonian with a constant and maximal norm $\|\hat{H}\| = J_0$.

\subsection{Cost functional}

The time-optimal dynamic is given by the extremum of the functional

\begin{eqnarray}
    \label{eq:action}
    S[\hat{U},\hat{\rho},\hat{R}] =\int_0^\tau (\mathcal{E}+\Tr(\hat{R}(\partial_t\hat{\rho}+i[\hat{H},\hat{\rho}])))dt\,,
\end{eqnarray}
where $\mathcal{E} = \frac{1}{2}\,\|\hat{H}\|^2+\Tr(\hat{D}\hat{H})$.
The operator 
$\hat{D}= \sum_{s}\lambda_{s}\hat{D}_s$
is introduced to impose restrictions on the sturcture of the Hamiltonian via Lagrange multipliers $\lambda_{s}$. Variation with respect to them yields the set of conditions
\begin{equation}
\text{Tr}(\hat{D}_s\hat{H}) = 0    
\end{equation}
valid at all moments of time. The second term in Eq.~\eqref{eq:action} prescribes unitary evolution of the density matrix $\hat{\rho} = \ket{\psi}\bra{\psi} = \hat{U}(t)\hat{\rho}_0\hat{U}^\dagger(t)$ via the Lagrange multiplier matrix $\hat{R}(t)$. As the operator $\partial_t\hat{\rho}+i[\hat{H},\hat{\rho}]$ is Hermitian, the matrix $\hat{R}$ can always be chosen Hermitian as well. Variation with respect to $\hat{R}$ yields von Neumann equation for the density matrix:
\begin{equation}\label{eq:Neumann}
\partial_t\hat{\rho}+i[\hat{H},\hat{\rho}]=0\:.    
\end{equation}

\subsection{Equations of motion}
%
Next we vary the cost functional with respect to the density matrix $\delta\hat{\rho}$
\begin{eqnarray}
    \delta S &=& \int_0^\tau dt\Tr\left(\hat{R}(\partial_t\delta\hat{\rho}+i[\hat{H},\delta\hat{\rho}])\right) = \int_0^\tau dt\Tr\left(\partial_t(\hat{R}\delta\hat{\rho})-\partial_t\hat{R}\,\delta\hat{\rho}+i[\hat{R},\hat{H}]\delta\hat{\rho})\right)\:.
\end{eqnarray}
This provides the equation of motion for $\hat{R}$
\begin{eqnarray}\label{eq:Revol}  \partial_t\hat{R}+i[\hat{H},\hat{R}] = 0\;.
\end{eqnarray}
Note that the initial and final density matrix $\hat{\rho}$ is fixed and thus $\delta\hat{\rho}(0) = \delta\hat{\rho}(\tau) = 0$.

Finally, we vary the functional with respect to the evolution operator $\delta \hat{U}$ by using the identity $\delta\hat{U}^\dagger=-\hat{U}^\dagger\,\delta\hat{U}\,\hat{U}^\dagger$ and the expression for the Hamiltonian in terms of the evolution operator:
\begin{equation}
\hat{H}=i \partial_t\hat{U} \,\hat{U}^\dagger=-i\hat{U}\,\partial_t\,\hat{U}^\dagger\:,
\end{equation}
which yields
\begin{eqnarray}
    \delta S &=& \int_0^\tau dt \left\lbrace\Tr\left(\hat{L} \delta\hat{H}\right) + \Tr\left(\hat{R}(i[\delta\hat{H},\rho])\right)\right\rbrace = \int_0^\tau dt \left\lbrace\Tr\left(\hat{L} \delta\hat{H}\right) + \Tr\left(\delta\hat{H}(i[\rho,\hat{R}])\right)\right\rbrace  \nonumber\\
    &=& \int_0^\tau dt \Tr\left((\hat{L} +i[\rho,\hat{R}])\delta\hat{H}\right)\equiv\int_0^\tau dt\,\Tr\left(\hat{F}\,\delta\hat{H}\right)\\\nonumber
    &=& i\,\int_0^\tau dt \Tr\left(
    \partial_t(\hat{F}\delta\hat{U}\hat{U}^\dagger)
    -\partial_t\hat{F}\delta\hat{U}\hat{U}^\dagger
    -i\left[\hat{H},\hat{F}\right]\delta\hat{U}\hat{U}^\dagger\right)\;,
\end{eqnarray}
where
\begin{equation}\label{eq:LaxDefinition}
\hat{L} = \frac{1}{2}\,\frac{\delta}{\delta\hat{H}}(\|\hat{H}\|^2)+\hat{D}    
\end{equation}
and $\hat{F} = \hat{L} +i[\rho,\hat{R}]$. Setting the variation $\delta S=0$, we immediately obtain the equations of motion
\begin{eqnarray}\label{eq:Fequation}
    \partial_t\hat{F} +i\left[\hat{H},\hat{F}\right] = 0\;,
\end{eqnarray}
with the boundary conditions
\begin{eqnarray}
    \hat{F}(0) = \hat{F}(\tau) = 0\;.
\end{eqnarray}
Equation~\eqref{eq:Fequation} has an immediate solution $\hat{F}(t)=\hat{U}(t)\,\hat{F}(0)\,\hat{U}^\dagger(t)$ and hence we conclude that $\hat{F}(t)=0$. Using the definition of $\hat{F}$, we recover
\begin{equation}\label{eq:Lbound}
    \hat{L}(t)=i[\hat{R}(t),\hat{\rho}(t)]\;.
\end{equation}
Since $\hat{R}(t)=\hat{U}(t)\hat{R}(0)\hat{U}^\dagger(t)$ and $\hat{\rho}(t)=\hat{U}(t)\,\hat{\rho}(0)\hat{U}^\dagger(t)$ due to Eqs.~\eqref{eq:Neumann},\eqref{eq:Revol}, the operator $\hat{L}$ follows the same evolution $\hat{L}(t)=\hat{U}(t)\hat{L}(0)\hat{U}^\dagger(t)$ due to Eq.~\eqref{eq:Lbound}. Hence, it satisfies the differential equation
\begin{eqnarray}\label{eqS:Lax}   
\partial_t\hat{L} +i\left[\hat{H},\hat{L}\right] = 0\;.
\end{eqnarray}

Equation~\eqref{eqS:Lax} is the celebrated quantum brachistochrone equation~\cite{Carlini2007} which defines the evolution of the Hamiltonian of an optimally controlled system. This equation should be supplemented by the boundary conditions obtained from Eq.~\eqref{eq:Lbound} at $t=0$ and $t=\tau$. Excluding an auxiliary matrix $\hat{R}$, we recover
\begin{eqnarray}
\hat{\rho}(0)\,\hat{L}(0)\,\hat{\rho}(0)=\left(\hat{I}-\hat{\rho}(0)\right)\,\hat{L}(0)\,\left(\hat{I}-\hat{\rho}(0)\right)=0\:,\label{eq:BC1}\\
\hat{\rho}(\tau)\,\hat{L}(\tau)\,\hat{\rho}(\tau)=\left(\hat{I}-\hat{\rho}(\tau)\right)\,\hat{L}(\tau)\,\left(\hat{I}-\hat{\rho}(\tau)\right)=0\:.\label{eq:BC2}
\end{eqnarray}
As the initial and final states $\hat{\rho}(0)$, $\hat{\rho}(\tau)$ are known, conditions Eqs.~\eqref{eq:BC1}-\eqref{eq:BC2} restrict the form of $\hat{L}$ operator at the initial and final moments of time.

The set of equations~\eqref{eqS:Lax},\eqref{eq:BC1},\eqref{eq:BC2} are the governing equations for the quantum brachistochrone approach. Importantly, we did not define the Hamiltonian norm so far, which allows one to apply this machinery for the various norm choices, i.e. various types of constraints.


\subsection{Lax operator}
The state of the quantum system evolves according to the Schr{\"o}dinger equation
\begin{equation}\label{eq:PsiEvol}
\partial_t\ket{\psi}=-i\hat{H}\,\ket{\psi}\:.
\end{equation}
Inspecting Eqs.~\eqref{eqS:Lax}, \eqref{eq:PsiEvol}, we recover that $-i\hat{H}$ and $\hat{L}$ operators form a Lax pair. As a consequence of that, the eigenvalues of the Lax operator $\hat{L}$ are time-independent~\cite{Babelon2003,stepanenko2025arxiv}. Since the density matrix $\rho = \ket{\psi}\bra{\psi}$, the commutator
\begin{eqnarray}
    \hat{L} = i[\hat{R},\rho] = i\left(\hat{R}\ket{\psi}\bra{\psi} - \ket{\psi}\bra{\psi}\hat{R}\right)
\end{eqnarray}
acts nontrivially only in the two-dimensional subspace spanned by
$\{|\psi\rangle, \hat{R}|\psi\rangle\}$, hence $\hat{L}$ has rank 2. Consequently, $\hat{L}$ has at most two nonzero eigenvalues, while the remaining $N-2$ eigenvalues vanish.

By using the Gram–Schmidt orthogonalization procedure, we recast the Lax operator as
\begin{eqnarray}
    \label{eqS:Lab}
    \hat{L} = L_0(\ket{a}\bra{a}-\ket{b}\bra{b})\;,
\end{eqnarray}
where  $\pm L_0$ are time-independent eigenvalues of $\hat{L}$, while corresponding eigenvectors are
\begin{eqnarray}
    \label{eqS:apsi}
    \ket{a} &=& \dfrac{1}{\sqrt{2}}(\ket{\psi}+\dfrac{i}{L_0}(\hat{R}\ket{\psi}-c\ket{\psi}))\;,\\
    \label{eqS:bpsi}
    \ket{b} &=& \dfrac{1}{\sqrt{2}}(\ket{\psi}-\dfrac{i}{L_0}(\hat{R}\ket{\psi}-c\ket{\psi}))\;.
\end{eqnarray}
The eigenvalue $L_0 = \sqrt{\bra{\psi}\hat{R}^2\ket{\psi}-\bra{\psi}\hat{R}\ket{\psi}^2}$ equals to uncertainty of $\hat{R}$, while $c = \bra{\psi}\hat{R}\ket{\psi}$.

If the dynamics of the vectors $\ket{a}$ and $\ket{b}$ is found, the wave function is restored from Eqs.~\eqref{eqS:apsi},\eqref{eqS:bpsi} as
\begin{equation}\label{eq:Psitoa}
\ket{\psi}=\frac{1}{\sqrt{2}}\,\left(\ket{a}+\ket{b}\right)\:.
\end{equation}
The elements of the Hamiltonian are recovered from Eq.~\eqref{eqS:Lab} by projecting $\hat{L}$ on the known basis matrices.

The above analysis shows that the entire problem can be parametrized by $2N+1$ quantities: the components $a_m =\bra{m}\ket{a}$ and $b_m =\bra{m}\ket{b}$, as well as $L_0$ constant. Such representation strongly reduces the number of variables from $\sim N^2$ to $\sim N$ and simplifies the solution of time-optimal control problem~\cite{stepanenko2025arxiv}.

\subsection{The norm and the penalty}

So far, we did not specify the choice of the norm in our problem. For the purposes of this study, we define the norm as
\begin{equation}
\|\hat{H}\|^2 = \Tr(\hat{G}(\hat{H})\hat{H})\:,    
\end{equation}
where the Hamiltonian reads
\begin{equation}\label{eqS:Hamiltonian} 
\hat{H}=\sum_{m=1}^{N-1}\sum_{p=1}^{N-m}J_{m,m+p}\left(i^{p-1}\ket{m}\bra{m+p}+i^{1-p}\ket{m+p}\bra{m}\right)
\end{equation}
and 
\begin{equation}
\hat{G}(\hat{H})=\sum_{m=1}^{N-1}\sum_{p=1}^{N-m}g_p\,J_{m,m+p}\left(i^{p-1}\ket{m}\bra{m+p}
+i^{1-p}\ket{m+p}\bra{m}\right)\:.  
\end{equation}
As a result,
\begin{equation}
\|\hat{H}\|^2 =2\,\sum\limits_{m=1}^{N-1}\,\sum\limits_{p=1}^{N-m}\,g_p\,J_{m,m+p}^2\:.
\end{equation}
Physically, the weights $g_p$ impose an additional penalty on the long-range couplings depending on the hopping distance $p$ and effectively suppress the maximal magnitude of such couplings.

With this choice of the norm, we calculate the Lax operator $\hat{L}$ by observing that $\Tr\left(\hat{G}(\delta\hat{H})\,\hat{H}\right)=\Tr\left(\hat{G}(\hat{H})\delta\hat{H}\right)$ and hence 
\begin{equation}\label{eq:LaxExpression}
\hat{L}\equiv\frac{1}{2}\,\frac{\delta}{\delta\hat{H}}\|\hat{H}\|^2+\hat{D}=\hat{G}(\hat{H})+\hat{D}\:.
\end{equation}

Therefore, the matrix elements of the Hamiltonian are computed in terms of the Lax operator as
%


%
\begin{eqnarray}
    H_{m,n} = L_{m,n}/g_{m,n}\;,
\end{eqnarray}
or using the equation~\eqref{eqS:Lab}
\begin{eqnarray}
    H_{m,n} = L_0(a_m a_n^*-b_m b_n^*)/g_{m,n}\;.
\end{eqnarray}

The evolution of the Lax eigenvector components is given by the projection of the Schrodinger equations
\begin{eqnarray}
    \label{eqS:sa}
    i\partial_t a_m &=& L_0\sum_{n}\dfrac{(a_m a_n^*-b_m b_n^*)}{g_{m,n}}a_n\;,\\
    \label{eqS:sb}
    i\partial_t b_m &=& L_0\sum_{n}\dfrac{(a_m a_n^*-b_m b_n^*)}{g_{m,n}}b_n\;.
\end{eqnarray}
with the boundary conditions derived from~\eqref{eqS:apsi}-\eqref{eqS:bpsi}.

\subsection{Chiral symmetry}

The Hamiltonian Eq.~\eqref{eqS:Hamiltonian} features an additional symmetry which establishes the connection between $\ket{a}$ and $\ket{b}$ vectors and allows further simplification of the solution. We introduce chiral symmetry operator $\hat{\Sigma}$ as
\begin{eqnarray}
    \hat{\Sigma} &=& \hat{Q}^2\hat{K}\;,\\
    \hat{Q} &=& \sum_m i^{m-1}\ket{m}\bra{m}\;,
\end{eqnarray}
where $\hat{K}$ is complex conjugation. Taken together, this yields
\begin{equation}
\hat{\Sigma}=\sum\limits_m\,(-1)^{m-1}\,\ket{m}\bra{m}\,\hat{K}\:.
\end{equation}
Note that $\hat{\Sigma}^2=\hat{I}$, i.e. $\hat{\Sigma}$ is a projective operator and $\hat{\Sigma}^{-1}=\hat{\Sigma}$.

Next we take an arbitrary matrix $\hat{B}$ from the space of $N\times N$ 
Hermitian matrices $\mathcal{M}$ and consider its transformation under chiral symmetry:
\begin{equation}\label{eqs:BC}
\hat{\Sigma}\,\hat{B}\,\hat{\Sigma}=\hat{C}\:,
\end{equation}
where we took into account that $\hat{\Sigma}^{-1}=\hat{\Sigma}$. Applying $\hat{\Sigma}$ to Eq.~\eqref{eqs:BC} once again, we recover
\begin{equation}\label{eqs:CB}
\hat{\Sigma}\,\hat{C}\,\hat{\Sigma}=\hat{B}\:.
\end{equation}
Combining Eqs.~\eqref{eqs:BC},\eqref{eqs:CB}, we derive:
\begin{gather}
\hat{\Sigma}\,\frac{1}{2}\,\left(\hat{B}+\hat{C}\right)\,\hat{\Sigma}=\frac{1}{2}\,\left(\hat{B}+\hat{C}\right)\:,\\
\hat{\Sigma}\,\frac{1}{2}\,\left(\hat{B}-\hat{C}\right)\,\hat{\Sigma}=-\frac{1}{2}\,\left(\hat{B}-\hat{C}\right)\:.
\end{gather}
Thus, an arbitrary operator $\hat{B}\in\mathcal{M}$ can be decomposed into the sum of the two parts: one anticommutes with $\hat{\Sigma}$ and belongs to $\mathcal{A}$ subspace, while another one commutes with $\hat{\Sigma}$ and belongs to $\mathcal{D}$ subspace.

The orthogonal basis matrices in these subspaces are chosen as follows:
\begin{eqnarray}
    \hat{A}_{m,n} &=& (i^{n-m-1}\ket{m}\bra{n}+i^{1-n+m}\ket{n}\bra{m})/\sqrt{2} \mspace{8mu}\text{for}\,m\neq n\;,\\
    \hat{D}_{m,n} &=& (i^{n-m}\ket{m}\bra{n}+i^{m-n}\ket{n}\bra{m})/\sqrt{2} \mspace{8mu}\text{for}\,m\neq n\;,\\
    \hat{D}_{m,m} &=& \left(\sum\limits_{k=1}^{m}\ket{k}\bra{k}-m\,\ket{m}\bra{m}\right)\,\sqrt{\frac{2}{m^2+m}}\mspace{8mu}\text{for}\, 1\leq m \leq N-1\:,\\
    \hat{D}_{N,N} &=& \hat{I}/\sqrt{N}\:.
\end{eqnarray}
%
%
Here, $\hat{A}_{m,n}$ anticommutes with $\hat{\Sigma}$, while $\hat{D}_{m,n}$ commutes with $\hat{\Sigma}$. Note that the Hamiltonian of the system $\hat{H}\in\mathcal{A}$. It is also straightforward to verify that 
\begin{align}
& i[\mathcal{A},\mathcal{A}]\in\mathcal{A}\:,\label{eqs:Com1}\\
& i[\mathcal{A},\mathcal{D}]\in\mathcal{D}\:,\label{eqs:Com2}\\
& i[\mathcal{D},\mathcal{D}]\in\mathcal{A}\:.\label{eqs:Com3}
\end{align}

Next we observe that the chosen initial density matrix $\hat{\rho}_0=\ket{1}\bra{1}$ commutes with the $\hat{\Sigma}$ operator and thus $\hat{\rho}\in\mathcal{D}$. Inspecting Eq.~\eqref{eq:Neumann} at the initial moment of time, when $\hat{\rho}\in\mathcal{D}$ and $\hat{H}\in\mathcal{A}$, we recover that $\partial_t\hat{\rho}\in\mathcal{D}$. Based on that, we conclude that the density matrix belongs to the $\mathcal{D}$ subspace at all moments of time:
\begin{equation}\label{eqs:rhod}
\hat{\rho}(t)\in\mathcal{D}.
\end{equation}
As a consequence, the final state of the system $\hat{\rho}_f\in\mathcal{D}$. If the final state is chosen differently, the solution to the quantum brachistochrone problem does not exist.

Next, we separate chiral and anti-chiral parts $\hat{R}_A$ and $\hat{R}_D$ of the Lagrange multiplier matrix $\hat{R}$. Equations~\eqref{eqs:Com1}-\eqref{eqs:Com3} then ensure that
\begin{align}
& \hat{L}_A=i\,[\hat{R}_D,\hat{\rho}]\:,\\
& \hat{L}_D=i\,[\hat{R}_A,\hat{\rho}]\:.
\end{align}
The structure of the quantum brachistochrone equation~\eqref{eqS:Lax} ensures that the dynamics of $\hat{L}_A$ and $\hat{L}_D$ parts decouples and thus we focus only on the dynamics of $\hat{L}_A$.
Similarly, to the previous subsection, $\hat{L}_A$ is a rank 2 Lax operator and takes form
\begin{eqnarray}
    \hat{L}_A = L_0(\ket{a}\bra{a}-\ket{b}\bra{b})\:,
\end{eqnarray}
where eigenvectors $\ket{a}$ and $\ket{b} = \hat{\Sigma}\ket{a}$. Further analysis is simplified if we use $\hat{Q}$ as a basis transformation.
In this basis, chiral symmetry operator acts as a complex conjugation $\hat{Q}\hat{\Sigma}\hat{Q}^{-1} = \hat{K}$ and thus $\ket{\beta} = \ket{\alpha^*}$, where $\ket{\alpha}=\hat{Q}\ket{a}$ and $\ket{\beta}=\hat{Q}\ket{b}$. All chiral matrices after transformation became imaginary 
\begin{eqnarray}
    \hat{H}^Q = \hat{Q}\hat{H}\hat{Q}^{-1} = -\hat{Q}\hat{H}^*\hat{Q}^{-1}\;,\\
    \hat{L}_A^Q = \hat{Q}\hat{L}_A\hat{Q}^{-1} = -\hat{Q}\hat{L}_A^*\hat{Q}^{-1}\;.
\end{eqnarray}

Having the Hamiltonian structure specified, we find coupling amplitudes
\begin{eqnarray}
    J_{m,n} = iL_0\dfrac{\alpha_m \alpha_{n}^* - \alpha_m^* \alpha_{n}}{g_{|m-n|}}
\end{eqnarray}

The system of governing equations~\eqref{eqS:sa}-\eqref{eqS:sb} takes the form
\begin{eqnarray}
    \label{eqS:sys_a}
    i\partial_t \alpha_m &=& L_0\sum_{n=1}^{N}\dfrac{\alpha_m \alpha_{n}^* - \alpha_m^* \alpha_{n}}{g_{|m-n|}} \alpha_n\;.
\end{eqnarray}

The boundary conditions are given by
\begin{eqnarray}
    \Re(\alpha_{m}(0)) = \delta_{1,m}/\sqrt{2}\;,\\
    \Re(\alpha_{m}(\tau)) = \delta_{N,m}/\sqrt{2}\;.
\end{eqnarray}

\section{Numerical solution}

From the numerical perspective, it is much simpler to solve a system of differential equations for real variables, especially in the nonlinear case. We introduce real and imaginary parts $\alpha_m = x_m + i y_m$ and equation~\eqref{eqS:sys_a} splits into
\begin{eqnarray}
    \partial_tx_m &=& \sum_n \dfrac{2L_0}{g_{m,n}}(y_{m}x_n - x_{m}y_n) x_n\;,\\
    \partial_t y_m &=& \sum_n \dfrac{2L_0}{g_{m,n}}(y_{m}x_n - x_{m}y_n) y_n\;.
\end{eqnarray}

The boundary conditions read
\begin{eqnarray}
    \ket{x(0)} = \ket{1}/\sqrt{2}\:,\\
    \ket{x(\tau)} = \ket{N}/\sqrt{2}\:.
\end{eqnarray}
To solve a boundary value problem, we use the shooting method. The obtained values of the transfer time $J_0\tau$ for various lengths of the array are provided in Fig.~3 of the main text and indicated in the Table~\ref{tab:1}.


\begin{table}[]
\begin{tabular}{|l|l|l|l|l|l|l|l|l|l|l|l|l|l|}
\cline{1-2} \cline{4-5} \cline{7-8} \cline{10-11} \cline{13-14}
$N$ & $J_0\tau$ &  & $N$ & $J_0\tau$ &  & $N$ & $J_0\tau$ &  & $N$ & $J_0\tau$ &  & $N$ & $J_0\tau$ \\ \cline{1-2} \cline{4-5} \cline{7-8} \cline{10-11} \cline{13-14} 
2 & $\pi/2$                 &  & 10 & 8.5840                  &  & 18 & 14.9936                 &  & 26 & 21.2386                 &  & 34 & 27.3955                 \\ 
3 & 2.5651                  &  & 11 & 9.3999                  &  & 19 & 15.7808                 &  & 27 & 22.0122                 &  & 35 & 28.1607                 \\ 
4 & 3.4885                  &  & 12 & 10.2104                 &  & 20 & 16.5659                 &  & 28 & 22.7845                 &  & 36 & 28.9250                 \\ 
5 & 4.3766                  &  & 13 & 11.0164                 &  & 21 & 17.3490                 &  & 29 & 23.5557                 &  & 37 & 29.6885                 \\ 
6 & 5.2427                  &  & 14 & 11.8183                 &  & 22 & 18.1301                 &  & 30 & 24.3257                 &  & 38 & 30.4513                 \\ 
7 & 6.0932                  &  & 15 & 12.6166                 &  & 23 & 18.9096                 &  & 31 & 25.0946                 &  & 39 & 31.2133                 \\ 
8 & 6.9321                  &  & 16 & 13.4117                 &  & 24 & 19.6874                 &  & 32 & 25.8625                 &  & 40 & 31.9746                 \\ 
9 & 7.7618                  &  & 17 & 14.2040                 &  & 25 & 20.4637                 &  & 33 & 26.6295                 &  & 41 & 31.9746                 \\ 
\cline{1-2} \cline{4-5} \cline{7-8} \cline{10-11} \cline{13-14} 
\end{tabular}
\caption{Calculated transfer times for the time-optimal protocol with the weight choice $g_p=p^2$.}
    \label{tab:1}
\end{table}

One of the most challenging aspects of numerical solution is to find an initial approximation for the vector $\ket{\alpha(0)}$. From the initial conditions we recover $x_m(0) = \mathrm{Re}(\alpha_m(0)) = \delta_{1m}/\sqrt{2}$, while the imaginary component $y_m(0) = \mathrm{Im}(\alpha_m(0))$ is unknown, with the only requirement $y_1(0)=0$.
To determine the unknown initial vector, we employ the shooting method implemented in Wolfram Mathematica. This approach requires an initial guess for $y_m(0)$, typically with an accuracy of about two decimal places to ensure convergence of the numerical procedure. For small systems (up to six qubits), relatively coarse guesses are sufficient to obtain the solution. These cases provide a useful intuition about the structure of the initial vector $\ket{y(0)}$.
For larger systems (up to approximately 10 qubits), we seek the initial approximation in the form
\begin{equation*}
  y_m(0)=-0.05 + \frac{1}{\sqrt{2 (N-m+1)}}\:,
\end{equation*}
%
where $N$ is the length of the qubit array, $m$ is the site number. In turn, the initial guess for $L_0$ is fitted by the $L_0 = 0.5\,N^{1.6} + 0.9 $ with the error below $0.5$ and can be extended for larger sizes. That allows us to construct a reasonable initial guess. However, this approach becomes inefficient for even larger systems.

To address this issue, we develop the following recurrent extrapolation procedure. First, all $3\leq n \leq N_{max}$ previously determined initial vectors $y_m^{(n)}(0)$,  are interpolated by a smooth function $f^{(n)}(q)$, where the upper index $(n)$ corresponds to the length of the chain and $q\in[2, n]$. They further discretized $f_z^{(n)}= f^{(n)}(z)$ to a common representation consisting of $N_z = 25$ equally distributed points $1\leq z\leq N_z$. The choice of $N_z = 25$ points is not critical but was found to be convenient for this analysis. Next, we assume that initial vectors for the system with the size $(n)$ and $(n+1)$ have only a minor difference, and thus $f^{(n)}_z$ is interpolated by a smooth function $F_z(n)$ of the size $n\in[3,N_{max}]$ for all $1\leq z\leq N_z$ of the array length $n$ for each index $z$. Using this assumption, we extrapolate the values $F_z(n)$ to $F_z(N_{max}+1)$ for $1\leq z\leq N_z$. This yields a 25-point approximation of the initial vector $f_z^{(N_{max}+1)} = F_z(N_{max}+1)$ for the $(N_{max}+1)$-qubit chain that needed further interpolation $f^{(N_{max}+1)}(z)$. Finally, this function $f^{(N_{max}+1)}(z)$ is descretized to $(N_{max}+1)$ equally distributed points $y_m^{(N_{max}+1)}(0) = f^{(N_{max}+1)}(m)$, which serve as the initial guess for $y_m(0)$ in the shooting method. 

The vector $y_m^{(N_{max}+1)}(0)$ serves as the initial guess for $y_m(0)$ in the shooting method. When a solution is found, we increase $N_{max}$ by one and repeat the procedure.


\begin{figure}
    \centering
    \includegraphics[width=0.5\linewidth]{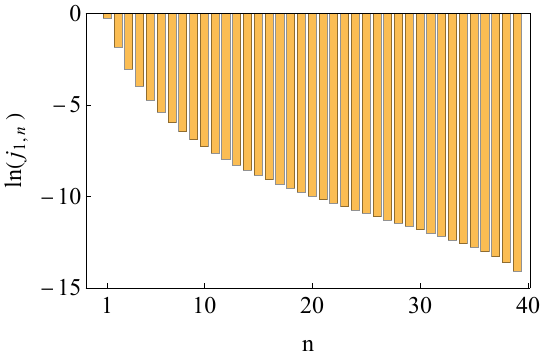}
    \caption{{\bf Probability currents for various trajectories in an optimally controlled lattice.}  Under the optimal protocol the state initially localized in the first qubit propagates by splitting into $N-1$ currents $j_{1,n} = 2J_{1,n}\psi_1^Q\psi_n^Q$  traversing all classical trajectories simultaneously. 
    }
    \label{fig:curr}
\end{figure}




\section{Transfer in an unpenalized qubit chain}

In this section, we consider a special case when all weights are $g_{m,n}=1$, so that there are no any specific restrictions on the long-range hoppings.

In such situation, the chiral part of the Lax operator $\hat{L} = \hat{H}$ due to Eq.~\eqref{eq:LaxExpression}. The Lax equation~\eqref{eqS:Lax} then ensures $\partial_t H = 0$. 

In addition, Lax eigenvectors become eigenvectors of the Hamiltonian $\hat{H}\ket{a}=L_0\ket{a} = J_0\ket{a}$. Since the Lax operator and Hamiltonian are constant, and their eigenvectors evolve only via global phase rotation
\begin{eqnarray}
    \ket{a(t)} &=& e^{-i\hat{H}t}\ket{a(0)} = e^{-iJ_0t}\ket{a(0)}\:.
\end{eqnarray}

The transition between states $\ket{\psi_0} = \ket{1}$ and 
$\ket{\psi_1} = \ket{N}$ is given by the function
\begin{eqnarray}
    \ket{\psi(t)} &=& (\ket{a}+\ket{a^*})/\sqrt{2} = (e^{-iJ_0t}\ket{a(0)} + e^{iJ_0t}\ket{a^*(0)})/\sqrt{2} \\
    &=& \sqrt{2}\cos(J_0 t)\ket{x(0)} +  \sqrt{2}\sin(J_0t)\ket{y(0)}\:.
\end{eqnarray}

The boundary conditions $\ket{\psi(0)} = \ket{\psi_0}$ and $\ket{\psi(\tau)} = \ket{\psi_1}$ yield
\begin{eqnarray}
    \ket{\psi_0} &=&\sqrt{2}\ket{x(0)}\;, \\
    \ket{\psi_1} &=&\sqrt{2}\cos(J_0 \tau)\ket{x(0)} +  \sqrt{2}\sin(J_0\tau)\ket{y(0)}\;.
\end{eqnarray}
Projecting the latter equation on $\bra{\psi_0}$, we find $\sqrt{2}\cos(J_0\tau)\bra{\psi_0}\ket{x(0)}=0$ due to $ \bra{x}\ket{y} = 0$, 
thus $J_0\tau = \pi (n+1/2)$ for $n\in \mathbb{N}$.

We thus conclude that time-optimal transfer in a lattice with no restrictions on the couplings is given by the constant couplings with only one non-zero amplitude $J_{1,N} = J_0$. The minimal time of the transfer $J_0\tau = \pi/2$ coincides with the lower bound given by the quantum speed limit for the orthogonal states~\cite{Deffner2017}. Therefore, the scenario with competing trajectories and quantum advantage is only possible provided the weights $g_p$ are different from unity. This is also clearly illustrated by the three-qubit example below.



\section{Three-qubit multi-path transfer protocol}
In this section, we present a 3-qubit system as the simplest and analytically solvable example where quantum advantage can be shown. We consider the transfer of a single-particle excitation localized in the first qubit to the last. The system of equations~\eqref{eqS:sys_a} is complicated to solve directly and therefore, we derive the equations for couplings:
\begin{eqnarray}\label{Jmmp}
    J_{m,m+p} &=& i H_{m,m+p}^Q = iL_{m,m+p}^Q/g_p = iL_0 (a_m a_{m+p}^* - a_m^* a_{m+p})/g_p\;
\end{eqnarray}
Calculating derivatives of couplings in this form and using the equation on Lax eigenvectors, we recover the differential equations on the couplings:
\begin{align}
    &\partial_t J_{1,2} = -(g-1)J_{2,3}J_{1,3}\:,\\
    &\partial_tJ_{2,3} = (g-1)J_{1,2}J_{1,3}\:,\\
    &\partial_tJ_{1,3} = 0\;.
\end{align}
Note that the same equations are obtained directly from quantum brachistochrone equation. The above equations have formal simple solutions:
\begin{eqnarray}
    J_{1,2} &=& A\cos(\Omega\,t + \varphi)\:,\\
    J_{2,3} &=& A\sin(\Omega\,t + \varphi)\:,\\
    J_{1,3} &=& B = \text{const}\:,
\end{eqnarray}
where $\Omega = (g-1)B$ and $A, B$, $\varphi$ are constants, determined from the boundary conditions. 

Moreover, because of the constant Hamiltonian norm: $J_{1,2}^2 + J_{2,3}^2 + g J_{1,3}^2 = J_0^2 = \text{const}$ we have relation: $A^2 + g\,B^2 = J_0^2$. We apply the boundary conditions on the wave function required by perfect transport: $\ket{\psi(0)} = (1, 0, 0)^T$ and $\ket{\psi(\tau)} = (0, 0, e^{i\phi})^T$, where $\tau$ is the transfer time. Also we use the connection between Lax eigenvectors $\ket{a}$ and $\ket{a^*}$ and wave function $\ket{\psi} = (\ket{a}+\ket{a^*})/\sqrt{2}$ to obtain boundary conditions on couplings: $J_{2,3} (0) = 0$ and $J_{1,2}(\tau) = 0$. These conditions yield $\varphi = 0$. Furthermore, these boundary conditions provide the result for the transfer time:
\begin{equation}\label{Ot}
    \Omega\,\tau = \frac{\pi}{2}+\pi n, \quad n\in\mathbb{Z}.
\end{equation}

To find other constants we need to construct solutions for wave function components from the Shr{\"o}dinger equation $\partial_t\ket{\psi} = \hat{H}\ket{\psi}$. Inspecting the equation for $\psi_2$ component, we find:
\begin{equation}
    \dddot{\psi}_2 + \omega^2\dot{\psi}_2=0,
\end{equation}
where $\omega^2 = A^2 + (\Omega + B)^2 = A^2 + g^2\,B^2$. It is a well-known equation for a harmonic oscillator with the solution:
\begin{equation}
    \psi_2 = C+D_+e^{-i\omega t} + D_-e^{i\omega t},
\end{equation}
where constants $C, D_+, D_-$ are determined from the  initial conditions. Using this and Shr{\"o}dinger equation, we find other components of the wave function:
\begin{eqnarray}
    \psi_1 &=& \frac{iCJ_{2,3}}{\Omega+B} + \frac{\omega J_{1,2} + i(\Omega + B)J_{2,3}}{A^2}D_+e^{-i\omega t} + \frac{-\omega J_{1,2} +i(\Omega +B)J_{2,3}}{A^2}D_-e^{i\omega t}\\
    \psi_3 &=& \frac{iCJ_{1,2}}{\Omega+B} + \frac{\omega J_{2,3} - i(\Omega + B)J_{1,2}}{A^2}D_+e^{-i\omega t} + \frac{-\omega J_{2,3} -i(\Omega +B)J_{1,2}}{A^2}D_-e^{i\omega t}
\end{eqnarray}
Next we apply the conditions $\psi_1(0)=1, \psi_2(0)=0, \psi_3(0)=0$ and $\psi_1(\tau)=0, \psi_2(\tau) = 0$, where $\tau$ is the transfer time. After applying these constrains we obtain the following constants: $C = 0, D_+=-D_- = \dfrac{A}{2\omega}$ and one more condition on transfer time:
\begin{equation}\label{ot}
    \omega\, \tau = \pi \, m, \quad m\in\mathbb{Z}.
\end{equation}
Then after some transformations from Eq.~\eqref{Ot} and Eq.~\eqref{ot} we recover the remaining constants:
\begin{eqnarray}
    A^2 &=& B^2\frac{4m^2(g-1)^2 - (1+2n)^2g^2}{(1+2n)^2}\\
    B^2 &=& \frac{J_0^2(1+2n)^2}{(g-1)(4m^2(g-1) - (1+2n)^2g)}\\
    \tau &=& \frac{\pi}{2J_0\sqrt{g-1}}\sqrt{4m^2(g-1) - (1+2n)^2g}
\end{eqnarray}
So, the last point is to find values for $m$ and $n$ to satisfy two conditions: minimum time and positive $A^2, B^2,$ and $\tau$. These requirements provide $m=1$ and $n=0,-1$. For these values, we recover the final answer:
\begin{eqnarray}
    \tau &=& \frac{\pi}{2J_0}\sqrt{\frac{3g-4}{g-1}},\\
    B &=& \frac{J_0}{\sqrt{(g-1)(3g-4)}},\\
    A &=& J_0\sqrt{\frac{(g-2)(3g-2)}{(g-1)(3g-4)}},
\end{eqnarray}
and from equation on $A$ we derive the restriction on $g \geq 2$. Finally, we recover the solutions, presented in the main text:
\begin{eqnarray}
    J_{1,2} &=& J_0\sqrt{\dfrac{(g-2)(3g-2)}{(g-1)(3g-4)}}\cos(\Omega\, t)\;,\\
    J_{2,3} &=& J_0\sqrt{\dfrac{(g-2)(3g-2)}{(g-1)(3g-4)}}\sin(\Omega\, t)\;,\\
    J_{1,3} &=& \frac{J_0}{\sqrt{(g-1)(3g-4)}}\;,
\end{eqnarray}
where $\Omega = (g-1)B=J_0\sqrt{(g-1)/(3g-4)}$.

\section{Estimate of transfer time for classical trajectory}
In this section, we estimate the transfer time for the set of classical trajectories to consistently show quantum advantage. By the classical trajectory we mean such hopping sequence, when the excitation starts in the first qubit, jumps from site to site in forward direction retaining localization, and eventually reaches the $N^{\text{th}}$ qubit. 

The only fixed points of such classical trajectory are first and $N^{\text{th}}$ qubits. For all other intermediate sites $m$ there are two possibilities: the excitation either visits this site or it skips the site by taking a longer-range hopping. Therefore, from simple combinatorics we calculate the total number of classical trajectories as $2^{N-2}$.


In case of short qubit arrays, we can compute the transfer time for all classical trajectories: the numerical results are presented in Table~S2, and illustrated as an inset in Fig.~3 of the main text. Note that the transfer time for our time-optimal solution is smaller than that for {\it any} of the classical trajectories, which indicates the onset of quantum advantage.

However, due to the exponential growth in the number of classical trajectories with the length of the array, it is not possible to repeat the above procedure for long qubit lattices. Therefore, we estimate a lower bound on the transfer time for all classical trajectories in the lattice of $N$ qubits. To that end, we consider a sequence of $f$ hoppings by $p_1$, $p_2$,\dots $p_f$ sites when the excitation visits the qubits $1+p_1$, $1+p_1+p_2$,\dots, $1+p_1+p_2+\dots +p_f$. By construction
\begin{equation*}
\sum\limits_{n=1}^f\,p_n=N-1\:.
\end{equation*}
The segment with hopping by $p_n$ sites takes not less than $\tau_s\,\sqrt{g(p_n)}$, where $p_n$ is the hopping distance and $\sqrt{g(p_n)}$ factor takes into account the slowdown of the excitation due to the constraint on the respective long-range coupling. $\tau_s$ is a constant factor which we specify below.

As a lower estimate for the transfer time, we sum the times of the consecutive segments choosing the weights $g(p)=p^2$:
\begin{equation}
\tau_{cl}=\sum\limits_{n=1}^{f}\,\tau_s\,\sqrt{g(p_n)}=\tau_s\,\sum\limits_{n=1}^{f}\,p_n=\tau_s\,(N-1)\:.
\end{equation}
In reality, the transfer takes longer, because the wavepacket could reshape during the propagation. Reshaping of the wave function close to the boundaries of the array takes an extra time as well.

The only remaining question is about the value of the $\tau_s$. We note that it could be taken from the optimal protocol for the single-particle excitation traveling in the nearest-neighbor-coupled qubit lattice~\cite{stepanenko2025arxiv}, where there is a single available classical trajectory~-- consecutive hopping between the qubits. The result reads: $J_0\,\tau_s=1.13031$. Thus, we finally arrive to the estimate
\begin{equation}\label{eq:classest}
J_0\,\tau_{cl}=1.13031\,(N-1)\:.
\end{equation}
If the transfer takes less than Eq.~\eqref{eq:classest}, this is an evidence that the excitation propagates non-classically, i.e. its propagation is not a sequence of hoppings. Note that the opposite is not necessarily true: the transfer may take more than Eq.~\eqref{eq:classest}, but still could be nonclassical.

An estimate Eq.~\eqref{eq:classest} is central to revealing quantum advantage in our system. By calculating the time-optimal protocol in the presence of constraints, we recover the transfer time substantially lower than prescribed by Eq.~\eqref{eq:classest}. This provides an immediate evidence of quantum advantage similarly to how Bell inequality violation is used to identify quantum entanglement.


\begin{table*}[b]
    \centering
    \begin{tabular}{|c|c|c|c|c|c|c|c|c|c|c|c|c|c|}
        \cline{1-14}
        $N$ & 3 & 3 & 4 & 4 & 4 & 5 & 5 & 5 & 5 & 5 & 5 & 6 & 6\\
        \cline{1-14}
        type & 1,2,3 & 1,3 & 1,2,3,4 & 1,2,4;1,3,4 & 1,4 & 1,2,3,4,5 & 1,2,3,5; 1,3,4,5 & 1,2,4,5 &  1,3,5 & 1,2,5; 1,4,5 & 1,5 & 1,3,6; 1,4,6& 1,6\\
        \cline{1-14}
        $J_0\tau$ & 2.72 & $\pi$ & 3.85 & 4.19 & $3\pi/2$ & 4.99& 5.30& 5.23&  5.44& 5.73& $2\pi$ & 6.87 & $5\pi/2$\\
        \hline
    \end{tabular}
\end{table*}
\begin{table}[h!]
    \centering
    \begin{tabular}{|c|c|c|c|c|c|c|c|c|c|}
        \hline
        $N$ & 6 & 6 & 6 & 6 & 6 & 6 & 6 & 6\\ 
        \cline{1-9}
         type & 1,2,3,4,5,6  &1,2,4,5,6;1,2,3,5,6& 1,2,3,4,6; 1,3,4,5,6 &  1,3,4,6 & 1,2,5,6 & 1,2,3,6; 1,4,5,6  & 1,2,4,6;1,3,5,6 & 1,2,6; 1,5,6\\ 
        \cline{1-9}
        $J_0\tau$ & 6.12 & 6.34 & 6.43 & 6.71 &6.75 & 6.84& 6.48 & 7.29\\ 
        \cline{1-9}
    \end{tabular}
    \caption{Transfer times $J_0\tau$ for the different classical trajectories in a qubit lattice with $N=3,4,5,6$ qubits. The type of the trajectory indicates the sequence of sites passed by the excitation. Weight choice $g_p=p^2$.}
    \label{tab:2}
\end{table}

\section{Asymptotic behavior of the transfer time in long lattices}

\begin{figure}[h]
    \centering
    \includegraphics[width=0.5\linewidth]{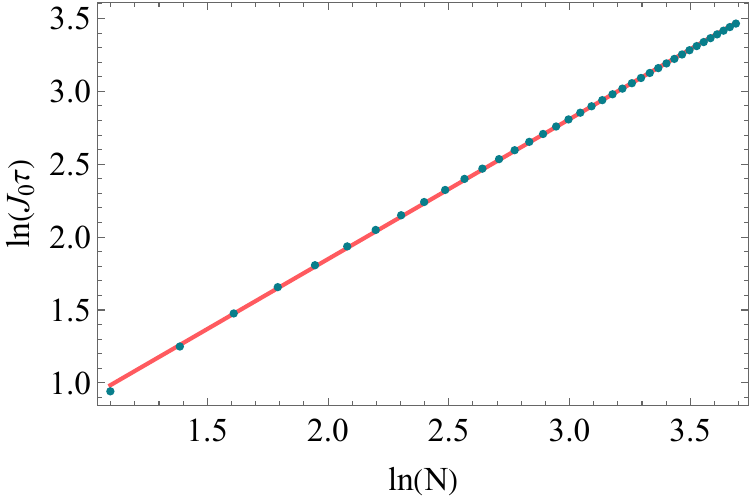}
    \caption{\textbf{Scaling of the transfer time with the length $N$ of the lattice.} Numerical results for the transfer time $J_0\tau$ versus lattice size $N$ in logarithmic scale (dots). Line shows the fit of the data  by the function $\ln(J_0\tau) = 0.960 \ln(N) - 0.071$ with an absolute error 0.042.}
    \label{fig:lntau}
\end{figure}

In this section, we examine the scaling of the transfer time with the length of the qubit lattice. First, we note, that for the small and intermediate-size lattices the time growth sub-linearly with $N$, which is clearly observed in double logarithmic scale in Fig.~\ref{fig:lntau}. This dependence can be fitted by the function $\ln(J_0\tau) = 0.960 \ln(N) - 0.071$ with an absolute error of 0.042. Thus, for $N\leq 40$ we find $J_0\tau \sim N^{z}$ with $z = 0.96$. 

However, with the increase of the lattice size $N$ the long-range current become more and more suppressed, and hence for larger sizes $N\rightarrow\infty$ we expect a linear dependence $\tau(N)$. To estimate the associated speed, we fit the numerically obtained data from the Table~\ref{tab:1} by the expansion $J_0\tau = 0.757 \,N + 2.018 - 14.198\, N^{-1} + 47.438\, N^{-2}-61.228\, N^{-3}$. Starting from the second term, all contributions in this expansion measure the boundary effects, wavepacket reshaping and other factors slowing down the excitation transfer. The absolute error of this expansion is 0.018. In the limit $N\rightarrow\infty$ negative powers due to boundary effects can be neglected yielding $J_0\tau = 0.757 \,N + 2.018$.

\section{The influence of constraints on the classical trajectory}

\begin{figure}
    \centering    \includegraphics[width=0.5\linewidth]{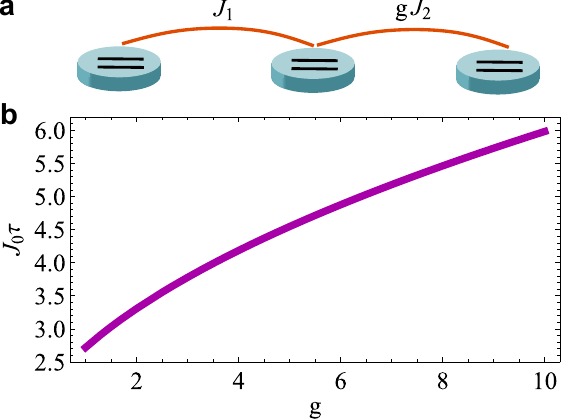}
    \caption{\textbf{Scheme of 3-qubit array and dependence of the transfer time on the weight $g$.} {\bf a} Scheme of a 3-qubit lattice with the nearest-neighbor couplings, where the second coupling is additionally restricted via weight $g$. {\bf b} The dependence of the transfer time on the magnitude of weight $g$. A monotonous increase is observed.}
    \label{fig:weights:time}
\end{figure}

In this section, we analyze the impact of constraints on the excitation transfer via single classical trajectory. As a representative example, we examine a lattice of three nearest-neighbor coupled qubits where the couplings satisfy the condition
\begin{equation}\label{eq:constraint3}
J_{1,2}^2+gJ_{2,3}^2 = J_0^2\:.    
\end{equation}
Physically this means that the maximal coupling between the second and the third qubits is suppressed compared to the first coupling [see Fig.~\ref{fig:weights:time}{\bf a}].
%
Using equation~\eqref{Jmmp}, we connect the couplings to the components of the Lax eigenvector:
\begin{eqnarray}
    J_{1,2} &=& iL_0 (a_1 a_{2}^* - a_1^* a_{2})\\
    J_{2,3} &=& i\frac{L_0}{g} (a_2 a_{3}^* - a_2^* a_{3}).
\end{eqnarray}
Using~\eqref{eqS:sys_a}, we derive the differential equations for the couplings:
\begin{eqnarray}
    \partial_t J_{1,2} &=& - \Omega\,J_{2,3}\\
    \partial_t J_{1,2} &=& \Omega\,J_{1,2}/g\\
    \partial_t \Omega &=& - (g-1)J_{1,2}J_{2,3},
\end{eqnarray}
where we introduced $\Omega = iL_0(a_1a_3^* - a_1^*a_3)$. Note that the same equations can be obtained directly from the quantum brachistochrone equation.

As we investigate the transfer of a single excitation from the first to the third qubit, the initial and boundary conditions read: $\ket{\psi(0)} = (1, 0, 0)^T$ and $\ket{\psi(\tau)} = (0, 0, -1)^T$, where $\tau$ is the transfer time, which yields $J_{2,3}(0) = 0$ and $J_{1,2}(\tau)=0$. 

Note that the structure of the above equations is similar to the differential equations for Jacobi elliptic functions.  Below, we provide a short summary of the properties of those functions relevant for our problem. Jacobi amplitude $\varphi\equiv \text{am}(u,m)$ is defined as the inverse of incomplete elliptic integral of the first kind:
%
\begin{equation}
     u = F(u,m) = \int_0^\varphi \frac{d\, \theta}{\sqrt{1-m\,\sin^2\theta}}.
\end{equation}
The elliptic functions are defined as:
\begin{eqnarray}
    \sn(u,m) &=& \sin\varphi\:,\\
    \cn(u,m) &=& \cos\varphi\:,\\
    \dn(u,m) &=& \frac{d\varphi}{d u} = \sqrt{1-m\sin^2\varphi}\:.
\end{eqnarray}
The zeros of the Jacobi sn function can be readily evaluated as
\begin{eqnarray}
    \sn(u,m) = 0 \quad \Rightarrow \quad u = \int_0^{\pi\,n} \frac{d\,\theta}{\sqrt{1-m\sin^2\theta}}=2n\int_0^{\pi/2} \frac{d\,\theta}{\sqrt{1-m\sin^2\theta}} = 2n\,K(m).
\end{eqnarray}
In a similar way the zeros of $\cn(u,m)$ are found at
$u_n=(2n+1)\,K(m)$ with the integer $n$. As it is straightforward to check, the elliptic functions defined this way satisfy the differential equations
\begin{eqnarray} 
    \partial_z \cn (z) &=& -\dn(z)\,\sn(z)\:,\\
    \partial_z \sn (z) &=& \dn(z)\,\cn(z)\:,\\
    \partial_z \dn (z) &=& -m\,\cn(z)\,\sn(z)\:.
\end{eqnarray}
Comparing these equations to the equations for the couplings above and considering the features of Jacobi elliptic functions, we recover the following solutions for the couplings:
\begin{eqnarray}
    J_{1,2} &=& J_0 \cn \left(J_0\, \omega \, t, \frac{g-1}{g\,\omega^2}\right)\\
    J_{2,3} &=& \frac{J_0}{\sqrt{g}} \sn \left(J_0\, \omega \, t, \frac{g-1}{g\,\omega^2}\right)\\
    \Omega &=& J_0\,\omega\,\sqrt{g} \dn \left(J_0\, \omega \, t, \frac{g-1}{g\,\omega^2}\right),
\end{eqnarray}
where $\omega$ is a constant determined by the initial and boundary conditions for the wave function $\ket{\psi}$, its numerical value can be found by the shooting method. Using the zeros of Jacobi $\cn$ function, we recover that the condition $J_{1,2}(\tau) =0$ is met when $\tau$ is equal to a quarter of the Jacobi elliptic function's period:
\begin{equation}
    J_0\tau_{g} = \frac{1}{\omega}\,K\left(\frac{g-1}{g\,\omega^2} \right),
\end{equation}
where $K(m)$ is the complete elliptic integral of the first kind. We illustrate the dependence of the couplings on time and the wave function evolution in Fig.~\ref{fig:weights} below.

Comparing the obtained transfer time with that in the absence of the weight $g$ [Fig.~\ref{fig:weights:time}{\bf b}], we recover an intuitive conclusion: adding any kind of constraints on the system necessarily increases the transfer time.

\begin{figure}
    \centering
    \includegraphics[width=1\linewidth]{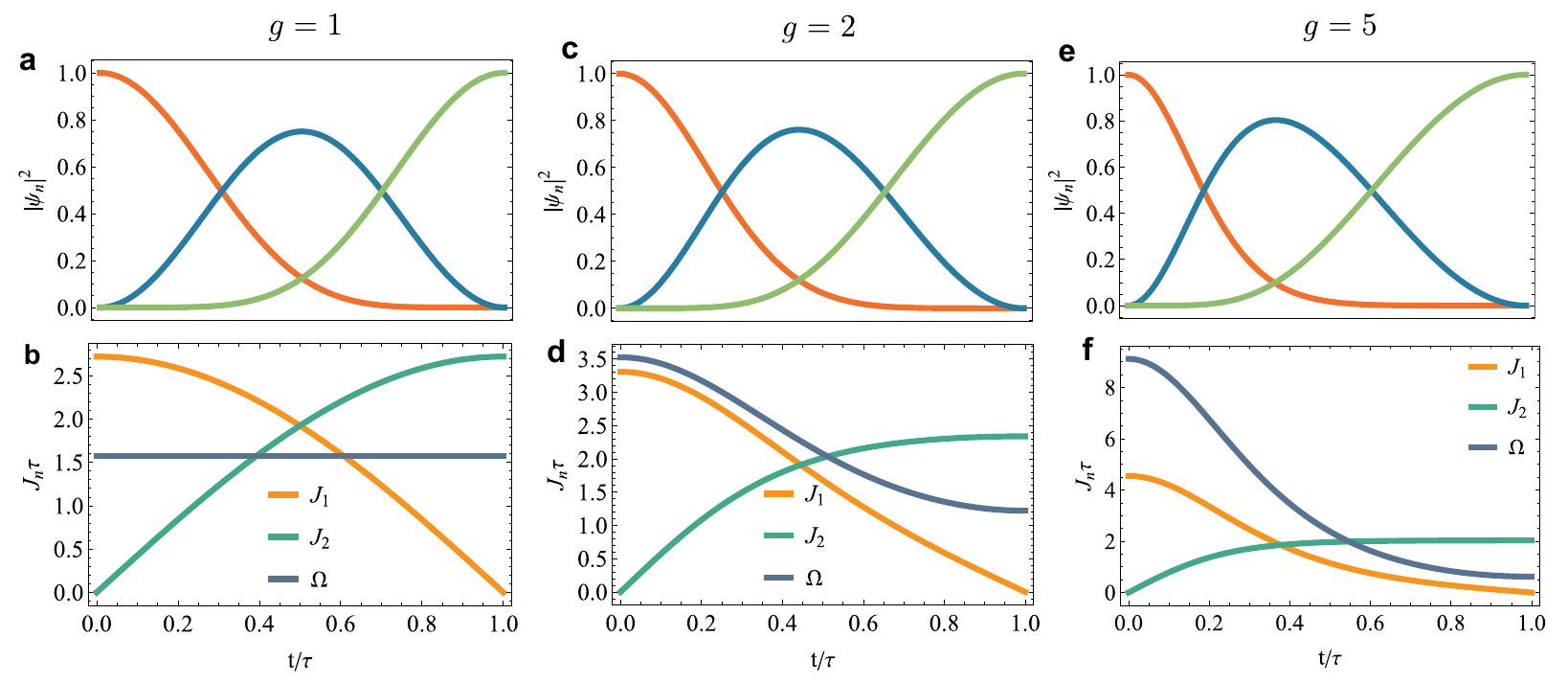}
    \caption{\textbf{Time-optimal transfer in the presence of additional constraints on the second coupling.} Evolution of the wave function (a,c,e) and the dependence of the couplings on time (b,d,f) for the time-optimal transfer protocol in a three-qubit lattice. The transfer protocol becomes markedly asymmetric when the weight $g$ is increased.}
    \label{fig:weights}
\end{figure}

\end{document}